\documentclass[twocolumn, times]{aastex631}

\graphicspath{{./}{figures/}}

\shorttitle{A Millisecond Pulsar Binary in the Galactic Center}
\shortauthors{Lower et al.}

\begin{document}

\title{A Millisecond Pulsar Binary Embedded in a Galactic Center Radio Filament}

\correspondingauthor{
    \\ Marcus~E.~Lower (\href{mailto: marcus.lower@csiro.au}{marcus.lower@csiro.au}),
    \\ Shi Dai (\href{mailto: shi.dai@csiro.au}{shi.dai@csiro.au}) \&
    \\ Simon Johnston (\href{mailto: simon.johnston@csiro.au}{simon.johnston@csiro.au}).
}

\author[0000-0001-9208-0009]{Marcus E. Lower}
\affiliation{Australia Telescope National Facility, CSIRO, Space and Astronomy, PO Box 76, Epping, NSW 1710, Australia}

\author[0000-0002-9618-2499]{Shi Dai}
\affiliation{Australia Telescope National Facility, CSIRO, Space and Astronomy, PO Box 76, Epping, NSW 1710, Australia}

\author[0000-0002-7122-4963]{Simon Johnston}
\affiliation{Australia Telescope National Facility, CSIRO, Space and Astronomy, PO Box 76, Epping, NSW 1710, Australia}

\author[0000-0001-8715-9628]{Ewan D. Barr}
\affiliation{Max-Planck-Institut f{\"u}r Radioastronomie, Auf dem H{\"u}gel 69, D-53121 Bonn, Germany}


\begin{abstract}
The Galactic Center is host to a population of extraordinary radio filaments, thin linear structures that trace out magnetic field lines running perpendicular to the Galactic plane.
Using Murriyang, the 64\,m Parkes radio telescope, we conducted a search for pulsars centered on the position of a compact source in the filament G359.0$-$0.2. 
We discovered a millisecond pulsar (MSP), PSR~J1744$-$2946, with a period $P = 8.4$\,ms, that is bound in a 4.8\,hr circular orbit around a $M_{\rm c} > 0.05\,M_{\odot}$ companion. 
The pulsar dispersion measure of $673.7 \pm 0.1$\,pc\,cm$^{-3}$ and Faraday rotation measure of $3011 \pm 3$\,rad\,m$^{-2}$ are the largest of any known MSP. 
Its radio pulses are moderately scattered due to multi-path propagation through the interstellar medium, with a scattering timescale of $0.87 \pm 0.08$\,ms at 2.6\,GHz.
Using MeerKAT, we localized the pulsar to a point source embedded in a low-luminosity radio filament, the ``Sunfish'', that is unrelated to G359.0$-$0.2.
Our discovery of the first MSP within 1\degr\ of the Galactic Center hints at a large population of these objects detectable via high frequency surveys. 
The association with a filament points to pulsars as the energy source responsible for illuminating the Galactic Center radio filaments.
\end{abstract}

\keywords{Binary pulsars (153) --- Galactic Center (565) --- Interstellar filaments (842) --- Pulsars (1306)} 

\section{Introduction} \label{sec:intro}
The quest to find pulsars in and around the Galactic Center has proved arduous, and mostly unsuccessful. 
To date, only a handful of long-period pulsars with high dispersion measures (DMs) have been found in blind searches \citep{jkl+06,dcl09} in addition to the magnetar SGR~1745$-$2900 \citep{Eatough2013, Rea2013}. 
Yet there is significant evidence for a substantial population of pulsars that reside in the Galactic Center, both from population analyses \citep{Pfahl2004,ploeg20} and as an explanation for the gamma-ray excess seen by the Fermi satellite \citep{OLeary2015, Bartels2016, Calore2021}. 
In addition, the Galactic Center hosts many supernova remnants and a plethora of other structures.
This includes the elongated non-thermal radio filaments which trace out magnetic field lines that run predominately perpendicular to the Galactic plane (see e.g the spectacular image in \citealt{Heywood2022}).
The exact origins of these filaments and the source of the energetic particles powering their emission is presently unknown, though links have been drawn between these structures and pulsars \citep{Barkov2019, Thomas2020}. 
The magnetic filaments themselves may be the result of cosmic-ray winds driven by activity within the Galactic Center (e.g., \citealt{Yusef-Zadeh2019}). 
This has since been reinforced through the discovery that many filaments appear to have compact radio sources embedded within them \citep{Yusef-Zadeh2022}.

Recently, \cite{Yusef-Zadeh2024} identified a point source (G359.13142$-$0.20005) surrounded by a diffuse structure and a tail-like feature in both Chandra X-ray and MeerKAT/Very Large Array radio images of the `major' kink in the non-thermal filament G359.1$-$0.2. 
Also known as the Snake, this filament is 70\,pc in length but less than 1\,pc wide, runs perpendicular to the Galactic plane and is located less than 1\degr\ from the Galactic Center \citep{gray91,Gray1995}.
\cite{Yusef-Zadeh2024} postulate that the point source is a neutron star with high spin-down energy ($\dot{E}$) embedded in a pulsar wind nebula (PWN) and that its fast motion (500-1000 kms$^{-1}$) through the dense medium produces the tail. 
In their picture, the neutron star could therefore have burst through the Snake sometime in the recent past causing the kink and (perhaps) energizing the Snake structure.

Motivated by the \cite{Yusef-Zadeh2024} paper we conducted a search for pulsars centered on their putative candidate.
While no radio pulses were detected from the position of G359.13142$-$0.20005, we discovered PSR~J1744$-$2946, the first millisecond pulsar to be found in the Galactic Center. 
We identify it as the likely source powering a non-thermal radio filament located $\sim$1\,arcmin to the west of the Snake, that we name the Sunfish.

\section{Observations} \label{sec:obs}

We conducted a target of opportunity observation (project code PX130) of G359.13142$-$0.20005 using the Murriyang Ultra-Wideband Low (UWL) receiver system \citep{Hobbs2020} on 2024 March 25. A total of 1338\,s of 2-bit, total intensity data were recorded using the {\sc medusa} signal processor, with 512\,$\mu$s sampling and 1\,MHz wide frequency channels covering the full 704-4032\,MHz band.
The resulting filterbank was then searched for periodic dispersed radio signals using the Fourier-domain acceleration search implemented in {\sc PRESTO}\footnote{\href{https://github.com/scottransom/presto}{https://github.com/scottransom/presto}} \citep{Ransom2011} after excising the data below 2.5\,GHz, similar to other UWL-based searches for highly scattered pulsars (e.g., \citealt{Lazarevic2023}).
We identified a pulsar candidate with a period $P = 8.4$\,ms and dispersion measure (DM) of $673$\,pc\,cm$^{-3}$.
A large drift in the pulsar period was apparent throughout the observation, indicating the presence of a binary companion.
Both the existence of the pulsar, named PSR~J1744$-$2946, and presence of a companion object were confirmed through additional follow-up observations where we recorded full Stokes search-mode data that were coherently dedispersed at the nominal DM of $673$\,pc\,cm$^{-3}$.
We also recorded separate calibration observations where the telescope pointing was offset from the pulsar and a switched noise source was injected into the signal chain.

We used the {\sc prepfold} function in {\sc PRESTO} to measure the pulsar spin period in 10-minute segments throughout each observation to determine the initial orbital parameters of the system.
The resulting spin period timeseries was then iteratively fit for a circular orbit using {\sc fit\_circular\_orbit.py}, which provided estimates of the orbital period, projected semi-major axis, epoch of the ascending node and de-Doppler shifted spin period.
We then used our initial timing solution to fold the search-mode data using {\sc dspsr} \citep{vanStraten2011}, averaging the data into 240\,s sub-integrations. 
The folded archives were then flux density and polarization calibrated and cleaned of radio-frequency interference (RFI) using {\sc psrchive}, following the same procedure as \cite{Lower2020}.

To facilitate a precise localization of the pulsar, we conducted a 1773\,s duration target of opportunity observation using the MeerKAT S-band receiver system and FBFUSE/APSUSE beamformer \citep{Barr2018}. 
We formed 54 tied-array beams overlapping at 70\,percent power that were placed on and around the locations of two radio point sources (27 beams each), G359.13142$-$0.20005 and G359.12460$-$0.18823. 
Both sources are within the 2.5 arcmin half-width half-maximum beam of Murriyang at 4\,Ghz.
Pulsar search data were recorded with 153\,$\mu$s sampling and 1024\,channels covering the S4 band (2625-3500\,MHz), which were then folded using {\sc dspsr} and excised of RFI using {\sc clfd} \citep{Morello2019}.

\begin{figure}
    \centering
    \includegraphics[width=\linewidth]{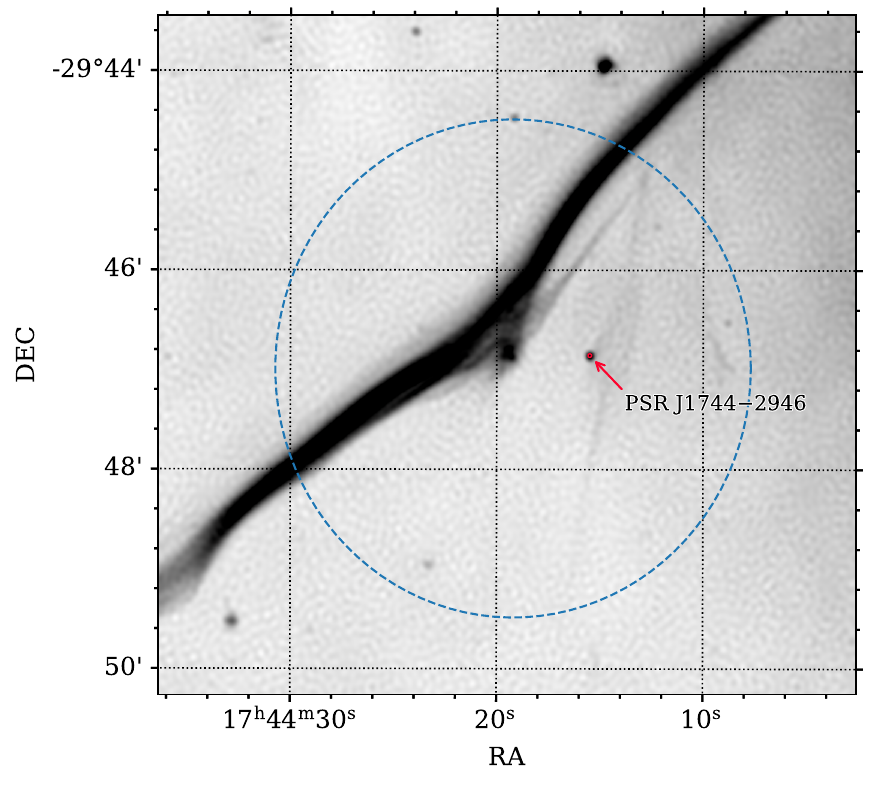}
    \caption{1.28\,GHz radio image of the region around the major kink in the Snake \citep{Heywood2022}. The $\sim$5\,arcmin diameter beam of Murriyang at 4\,GHz is shown by the large blue circle centered on G359.13142$-$0.20005. The red circle is centered on the MeerKAT S-band localization of PSR~J1744$-$2946.}
    \label{fig:image}
\end{figure}

\section{Results} \label{sec:res}
PSR~J1744$-$2946 was only detected in beams that were placed on and around G359.12460$-$0.18823, the source that is not associated with the Snake.
Using the {\sc Mosaic}\footnote{\href{https://github.com/wchenastro/Mosaic}{https://github.com/wchenastro/Mosaic}} and {\sc SeeKAT}\footnote{\href{https://github.com/BezuidenhoutMC/SeeKAT}{https://github.com/BezuidenhoutMC/SeeKAT}} packages \citep{Bezuidenhout2023}, we obtained a sub-arcsecond precision localization of the pulsar (see Table~\ref{tab:params}).
A comparison between the 4\,GHz Murriyang beam and MeerKAT S-band localization of PSR~J1744$-$2946 is shown in Figure~\ref{fig:image} overlaid against the 1.28\,GHz Galactic Center image of \cite{Heywood2022}. 
Its position is consistent with the point source G359.12460$-$0.18823, which is itself embedded within a filamentary structure approximately 1\,arcmin west of the Snake.
The overall shape of this filament resembles a sunfish.

\begin{figure}
    \centering
    \includegraphics[width=\linewidth]{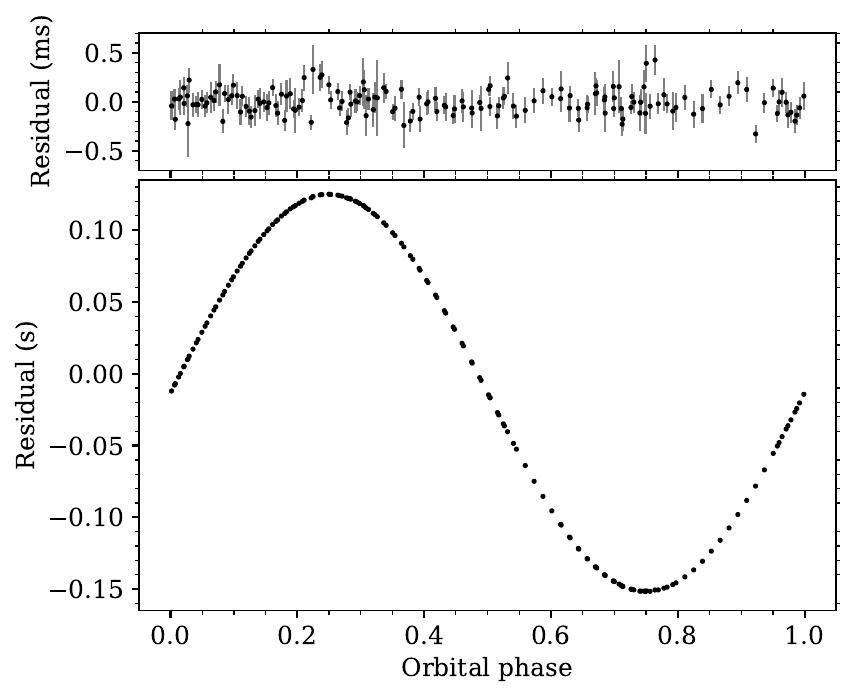}
    \caption{Top panel shows the pulsar timing residuals of PSR~J1744$-$2946 as a function of orbital phase after applying the timing model in Table~\ref{tab:params}. Bottom panel assumes a binary semi-major axis of zero to demonstrate the influence of the companion object.}
    \label{fig:timing}
\end{figure}

We used pulsar timing to refine our measurements of the pulsar properties. 
Pulse times of arrival (ToAs) were measured by cross-correlating the frequency-averaged profiles with a noiseless template, after which we used {\sc tempo2} to re-fit the pulsar spin frequency and orbital parameters.
The pulsar is in a circular orbit with a period of 4.8~hr about a companion of mass $> 0.05\,M_{\odot}$. 
Details of the pulsar and binary parameters are given in Table~\ref{tab:params}.
Figure~\ref{fig:timing} shows the timing residuals both with and without the full timing model presented in Table~\ref{tab:params} being subtracted from the ToAs.
A small delay is apparent in the ToAs near orbital phase 0.25, which could be due to interactions with material being ablated off the companion.
However, we find no evidence of radio eclipses or flux density variability at any part of the orbit.

\begin{table}
\begin{center}
\caption{Measured and derived parameters for PSR~J1744$-$2946.}
\label{tab:params}
\begin{tabular}{ll}
\hline
\hline
Pulsar parameters &  \\
\hline
Right ascension, R.A. (J2000)$^{\dagger}$              & 17$^{\mathrm{h}}$44$^{\mathrm{m}}$15.502(8)$^{\mathrm{s}}$ \\
Declination, Decl. (J2000)$^{\dagger}$                 & $-$29$^{\circ}$46$'$51.65(5)$''$ \\
Spin frequency, $\nu$ (Hz)                             & $119.158395816(2)$ \\
Epoch (MJD TCB)                                        & 60394 \\
Dispersion measure, DM (pc\,cm$^{-3}$)                 & $673.7(1)$ \\
Rotation measure, RM (rad\,m$^{-2}$)                   & $3011(3)$ \\
Solar System ephemeris                                 & DE438 \\
Time span (MJD)                                        & 60394-60403 \\
Number of ToAs                                         & $167$ \\
Flux density at 2.1\,GHz, $S_{2.1}$ (mJy)              & $0.43(4)$ \\
Scattering timescale at 2.6\,GHz (ms)                  & $0.87(8)$ \\
Pulse width, $W_{10}$ (ms)                             & $6.0(1)$ \\
\hline
Binary model & ELL1 \\
Projected semi-major axis, $x$ (lt-s)                  & $0.13821(1)$ \\
Epoch of ascending node, $T_{0}$ (MJD)                 & $60394.055440(7)$ \\
Orbital period, $P_{b}$ (d)                            & $0.19918615(9)$ \\
Eccentricity, $e$                                      & $\lesssim 0.0003$ \\
\hline
Spin period, $P$ (ms)                                  & $8.3921908578(1)$ \\
Distance, $d$ (kpc)$^{\star}$                          & 8.4 \\
Radio luminosity, $S_{2} d^{2}$ (mJy\,kpc$^{2}$)       & 30(1) \\
Mass function, $f(M_{p})$ $(M_{\odot})$                & $0.00007145(2)$ \\
Minimum companion mass, $M_{\rm c}^{\rm min}$ $(M_{\odot})$ & $0.0520$ \\
Median companion mass, $M_{\rm c}^{\rm med}$ $(M_{\odot})$  & $0.0603$ \\
\hline\\
\end{tabular}
\end{center}
\vspace{-0.5cm}
{\bf Notes.} Values in parentheses represent 1$\sigma$ uncertainties. {$^{\dagger}$}Position from {\sc SeeKAT}. {$^{\star}$}Distance inferred from the DM and the \citet{Cordes2002} model.
\end{table}

\begin{figure}
    \centering
    \includegraphics[width=\linewidth]{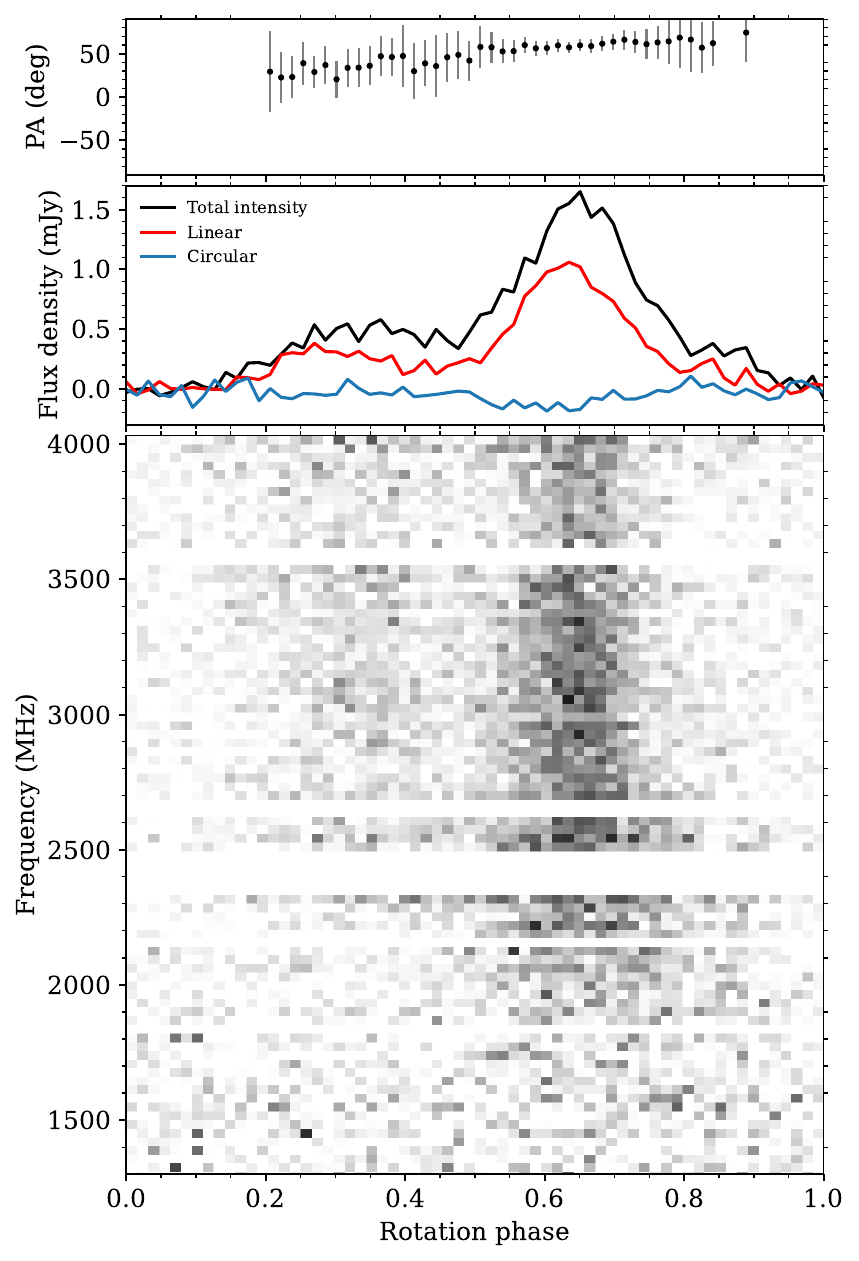}
    \caption{Linear polarization position angle swing (top), time and frequency averaged polarization profile (middle) and the total intensity spectrum (bottom) of PSR~J1744$-$2946. Note, the average profile in the middle panel excluded data below 1900\,MHz.}
    \label{fig:profile}
\end{figure}

Spectropolarimetric analysis was conducted using {\sc psrchive} \citep{Hotan2004, vanStraten2011} where a Faraday rotation measure (RM) of $3011 \pm 3$\,rad\,m$^{-2}$ was recovered.
This is the sixth largest absolute RM of any pulsar found to date.
In Figure~\ref{fig:profile}, we present both the polarization profile of PSR~J1744$-$2946 and its rotational phase resolved radio spectrum after averaging together the observations taken on 2024 April 1 and 3.
Its average profile is highly polarized with a linear fraction of $58$\,percent, with a largely flat position angle sweep across the profile.

We obtained a refined DM by splitting the data between 2880-4032\,MHz into 128\,MHz subbands and then fitting the resulting multi-frequency ToAs using {\sc tempo2}, yielding an improved ${\rm DM} = 673.7 \pm 0.1$\,pc\,cm$^{-3}$.
The pulsar is scatter broadened by multi-path radio wave propagation in the interstellar medium, to the point where it is not detected below $\sim$1500\,MHz. 
Fitting the profile at 2.6\,GHz with a noise-free template generated from the 4\,GHz profile convolved with an exponential tail returned a scattering timescale of $0.87 \pm 0.08$\,ms. Under a $f^{-4}$ frequency scaling relation, this would correspond to a scattering timescale of $\sim$40\,ms at 1\,GHz, less than a factor 2 from the scattering-DM relationship discussed in \citet{coc22}.

Flux densities are difficult to obtain due to both the effects of scatter-broadening and the wide pulse profile. 
We divided the band into 14 subbands and measured the flux density in each band by simply summing the values across all phase-bins. The lowest bands have flux densities consistent with zero.
Figure~\ref{fig:flux} shows the phase-averaged flux density of the remaining subbands as a function of observing frequency along with continuum flux densities at 1.28, 6.0 and 10\,GHz obtained from MeerKAT \citep{Heywood2022} and Very Large Array imaging data \citep{Yusef-Zadeh2024}. 
A clear turnover in the Murriyang values below 2.2\,GHz results from pulse scatter broadening.
We find pulsar spectrum to be well described by as simple power-law fit with a spectral index of $-1.5 \pm 0.1$ and a 1.4~GHz flux density of 1.6~mJy.

\section{Discussion} \label{sec:disc}
Our localization of PSR~J1744$-$2946 to the Sunfish using MeerKAT rules out an association between it and the Snake.
A putative pulsar powering G359.13142$-$0.20005 remains undetected. From our Murriyang observations we can set an upper limit on the 2.5-4\,GHz flux density of $0.03$\,mJy assuming a pulsar with a 20\,percent pulse duty cycle.
This value is well below the continuum flux density reported by \citet{Yusef-Zadeh2024} and suggests that a pulsar in the Snake, if it exists, must be highly scatter broadened even at 4\,GHz.

The DM and RM of PSR~J1744$-$4429 are the largest of any MSP discovered to date, and point to it residing in the central region of our Galaxy. 
For a DM of 673.3~cm$^{-3}$pc in the direction of the Snake, the YMW16 \citep{Yao2017} and NE2001 \citep{Cordes2002} electron column density models for the Galaxy yield distance estimates of 4.6 and 8.4\,kpc respectively. 
While DM distances should be treated with caution, particularly for the inner Galaxy, the larger NE2001 distance is consistent with a lower-bound of $\sim$8\,kpc to the Snake inferred via {\sc Hi} absorption \citep{umy92}. 
The DM of PSR~J1744$-$2946 is noticeably lower than those of other pulsars in the vicinity of the Galactic Center which are typically $>1000$~cm$^{-3}$pc. 
While the high RM is an order of magnitude lower than those of several slow pulsars in the Galactic Center \citep{sef+16,sj13}, it aligns well with the RM measured in the Snake itself \citep{Gray1995} and that of the filament G359.54$+$0.18 \citep{Yusef-Zadeh1997}.
These other pulsars also reside much closer to Sagittarius A ($<0.5\degr$ on the sky), hence the comparatively lower DM and RM of PSR~J1744$-$2946 may not be too surprising given a similar distance. 

\begin{figure}
    \centering
    \includegraphics[width=\linewidth]{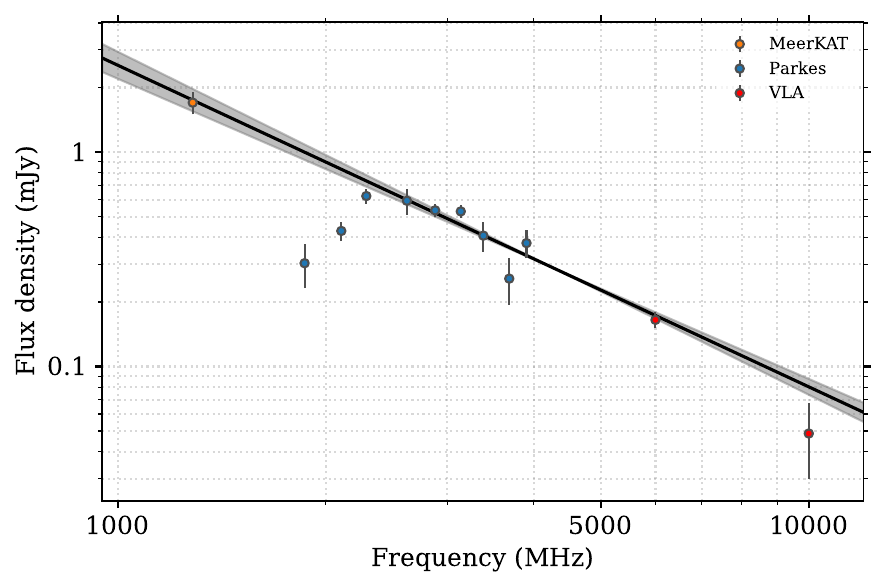}
    \caption{Flux density of PSR~J1744$-$4429 as a function of observing frequency. Over-plotted in black is our best-fit spectral model and 68\,percent confidence interval (gray shading).}
    \label{fig:flux}
\end{figure}

The pulsar has a very broad profile which appears to fill most of the 360\degr\ of rotational longitude. While many MSPs have broad profiles \citep{spiewak22} these are usually composed of many different separate components rather than the two co-joined components seen here. The high degree of linear polarization is typical of MSPs with high $\dot{E}$ values.

In Figure~\ref{fig:flux} we show the power-law spectrum of PSR~J1744$-$2946, where its spectral index of $-1.5 \pm 0.1$ is quite typical of radio pulsars \citep{jvk+18}.
At a distance of 8.4\,kpc the 1.4\,GHz luminosity is $\sim$113\,mJy\,kpc$^{2}$, making PSR~J1744$-$2946 the second most luminous MSP in the Galaxy at centimeter wavelengths after PSR~B1937$+$21.
This unusually high radio luminosity may point to it being the brightest member of an as-of-yet undetected population of lower-luminosity MSPs residing in the Galactic Center.
However, there are no known gamma-ray MSPs with spin periods as long as 8.4~ms \citep{Smith2023}, so it would appear that the pulsar is not a contributor to the gamma-ray excess seen towards the Galactic Center.

The morphology of the Sunfish is remarkably similar to that of the neighboring Snake.
A diffuse triangular wedge of radio emission appears to connect the pulsar to a larger, linear structure.
The potentially low $\dot{E}$ of PSR~J1744$-$2946 fits with the lack of an X-ray counterpart (see Figure 3 of \citealt{Yusef-Zadeh2024}) along with the low surface brightness of the Sunfish given the expected scaling between filament luminosity and pulsar $\dot{E}$ \citep{Barkov2019}.
As suspected for the major kink in the Snake \citep{Yusef-Zadeh2024}, the wedge-shaped feature may be the result of PSR~J1744$-$2946 having punched through the Sunfish. 
Its orientation could then indicate the pulsars direction of motion.
For a typical MSP transverse velocity of $\sim$100\,km\,s$^{-1}$ and a distance of 8.4\,kpc, the pulsar would only take $\sim$7.6\,kyr to traverse the 19\,arcsec distance from the linear part of the Sunfish to its current position. 
This is far smaller than the expected lifetime of a MSP, providing ample time for it to inject energy into the filament.
The association between PSR~J1744$-$2946 and the Sunfish will ultimately be tested through future measurements of its proper motion and parallax distance obtained via very-long baseline interferometry. 

In summary, we have discovered the first MSP in the Galactic Center. In spite of the pulsar’s relatively high radio luminosity, scattering in the interstellar medium renders its pulses invisible below 1.8\,GHz. 
Future surveys of the entire Galactic Center region at 3\,GHz and above are therefore warranted. 
Detection of a large population of MSPs would lend support to the idea that the Fermi GeV excess in this region arises from such a population \citep{Calore2016}.
Its association with the Sunfish, alongside the putative pulsar in the Snake \citep{Yusef-Zadeh2024}, presents a compelling case that pulsars are the energy source responsible for energizing the Galactic Center radio filaments. 
A near-future measurement of the $\dot{P}$ of PSR~J1744$-$2946 will provide a measure of the efficiency at which pulsar rotational energy is transferred to these structures.

\begin{acknowledgments}
We are grateful to the CSIRO staff, especially Jane Kaczmarek and Lawrence Toomey, for helping support the Murriyang observations, and to Sarah Buchner and the SARAO staff for scheduling and conducting the MeerKAT observation.
We thank Saurav Mishra for suggestions on pulsar searching and Juntao Bai for the code to compute scattering times.
We also thank Farhad Yusef-Zadeh for supplying us with the Very Large Array C-band and X-band images. 
Murriyang, the Parkes radio telescope, is part of the Australia Telescope National Facility (\href{https://ror.org/05qajvd42}{https://ror.org/05qajvd42}) which is funded by the Australian Government for operation as a National Facility managed by CSIRO.
We acknowledge the Wiradjuri people as the traditional owners of the Observatory site.
The MeerKAT telescope is operated by the South African Radio Astronomy Observatory (SARAO), which is a facility of the National Research Foundation, an agency of the Department of Science and Innovation. 
SARAO acknowledges the ongoing advice and calibration of GPS systems by the National Metrology Institute of South Africa (NMISA) and the time space reference systems department department of the Paris Observatory. 
This work made use of the FBFUSE/APSUSE computing cluster which is funded and operated by the Max-Planck-Institut f{\"u}r Radioastronomie and the Max-Planck-Gesellschaft.
This project was supported by resources and expertise provided by CSIRO IMT Scientific Computing.
\end{acknowledgments}

\vspace{5mm}
\facilities{Parkes/Murriyang, MeerKAT}

\software{
{\sc clfd}~\citep{Morello2019},
{\sc dspsr}~\citep{vanStraten2011},
{\sc Mosaic}~\citep{Bezuidenhout2023},
{\sc PRESTO}~\citep{Ransom2011},
{\sc psrchive}~\citep{Hotan2004, vanStraten2012},
{\sc SeeKAT}~\citep{Bezuidenhout2023},
{\sc tempo2}~\citep{Hobbs2006}
          }

\bibliography{main}{}

\begin{thebibliography}{}
\expandafter\ifx\csname natexlab\endcsname\relax\def\natexlab#1{#1}\fi
\providecommand{\url}[1]{\href{#1}{#1}}
\providecommand{\dodoi}[1]{doi:~\href{http://doi.org/#1}{\nolinkurl{#1}}}
\providecommand{\doeprint}[1]{\href{http://ascl.net/#1}{\nolinkurl{http://ascl.net/#1}}}
\providecommand{\doarXiv}[1]{\href{https://arxiv.org/abs/#1}{\nolinkurl{https://arxiv.org/abs/#1}}}

\bibitem[{{Barkov} \& {Lyutikov}(2019)}]{Barkov2019}
{Barkov}, M.~V., \& {Lyutikov}, M. 2019, \mnras, 489, L28,
  \dodoi{10.1093/mnrasl/slz124}

\bibitem[{{Barr}(2018)}]{Barr2018}
{Barr}, E.~D. 2018, in Pulsar Astrophysics the Next Fifty Years, ed.
  P.~{Weltevrede}, B.~B.~P. {Perera}, L.~L. {Preston}, \& S.~{Sanidas}, Vol.
  337, 175--178, \dodoi{10.1017/S1743921317009036}

\bibitem[{{Bartels} {et~al.}(2016){Bartels}, {Krishnamurthy}, \&
  {Weniger}}]{Bartels2016}
{Bartels}, R., {Krishnamurthy}, S., \& {Weniger}, C. 2016, \prl, 116, 051102,
  \dodoi{10.1103/PhysRevLett.116.051102}

\bibitem[{{Bezuidenhout} {et~al.}(2023){Bezuidenhout}, {Clark}, {Breton},
  {Stappers}, {Barr}, {Caleb}, {Chen}, {Jankowski}, {Kramer}, {Rajwade}, \&
  {Surnis}}]{Bezuidenhout2023}
{Bezuidenhout}, M.~C., {Clark}, C.~J., {Breton}, R.~P., {et~al.} 2023, RAS
  Techniques and Instruments, 2, 114, \dodoi{10.1093/rasti/rzad007}

\bibitem[{{Calore} {et~al.}(2016){Calore}, {Di Mauro}, {Donato}, {Hessels}, \&
  {Weniger}}]{Calore2016}
{Calore}, F., {Di Mauro}, M., {Donato}, F., {Hessels}, J.~W.~T., \& {Weniger},
  C. 2016, \apj, 827, 143, \dodoi{10.3847/0004-637X/827/2/143}

\bibitem[{{Calore} {et~al.}(2021){Calore}, {Donato}, \& {Manconi}}]{Calore2021}
{Calore}, F., {Donato}, F., \& {Manconi}, S. 2021, \prl, 127, 161102,
  \dodoi{10.1103/PhysRevLett.127.161102}

\bibitem[{{Cordes} \& {Lazio}(2002)}]{Cordes2002}
{Cordes}, J.~M., \& {Lazio}, T.~J.~W. 2002, arXiv e-prints, astro,
  \dodoi{10.48550/arXiv.astro-ph/0207156}

\bibitem[{{Cordes} {et~al.}(2022){Cordes}, {Ocker}, \& {Chatterjee}}]{coc22}
{Cordes}, J.~M., {Ocker}, S.~K., \& {Chatterjee}, S. 2022, \apj, 931, 88,
  \dodoi{10.3847/1538-4357/ac6873}

\bibitem[{{Deneva} {et~al.}(2009){Deneva}, {Cordes}, \& {Lazio}}]{dcl09}
{Deneva}, J.~S., {Cordes}, J.~M., \& {Lazio}, T.~J.~W. 2009, \apjl, 702, L177,
  \dodoi{10.1088/0004-637X/702/2/L177}

\bibitem[{{Eatough} {et~al.}(2013){Eatough}, {Falcke}, {Karuppusamy}, {Lee},
  {Champion}, {Keane}, {Desvignes}, {Schnitzeler}, {Spitler}, {Kramer},
  {Klein}, {Bassa}, {Bower}, {Brunthaler}, {Cognard}, {Deller}, {Demorest},
  {Freire}, {Kraus}, {Lyne}, {Noutsos}, {Stappers}, \& {Wex}}]{Eatough2013}
{Eatough}, R.~P., {Falcke}, H., {Karuppusamy}, R., {et~al.} 2013, \nat, 501,
  391, \dodoi{10.1038/nature12499}

\bibitem[{{Gray} {et~al.}(1991){Gray}, {Cram}, {Ekers}, \& {Goss}}]{gray91}
{Gray}, A.~D., {Cram}, L.~E., {Ekers}, R.~D., \& {Goss}, W.~M. 1991, \nat, 353,
  237, \dodoi{10.1038/353237a0}

\bibitem[{{Gray} {et~al.}(1995){Gray}, {Nicholls}, {Ekers}, \&
  {Cram}}]{Gray1995}
{Gray}, A.~D., {Nicholls}, J., {Ekers}, R.~D., \& {Cram}, L.~E. 1995, \apj,
  448, 164, \dodoi{10.1086/175949}

\bibitem[{{Heywood} {et~al.}(2022){Heywood}, {Rammala}, {Camilo}, {Cotton},
  {Yusef-Zadeh}, {Abbott}, {Adam}, {Adams}, {Aldera}, {Asad}, {Bauermeister},
  {Bennett}, {Bester}, {Bode}, {Botha}, {Botha}, {Brederode}, {Buchner},
  {Burger}, {Cheetham}, {de Villiers}, {Dikgale-Mahlakoana}, {du Toit},
  {Esterhuyse}, {Fanaroff}, {February}, {Fourie}, {Frank}, {Gamatham}, {Geyer},
  {Goedhart}, {Gouws}, {Gumede}, {Hlakola}, {Hokwana}, {Hoosen}, {Horrell},
  {Hugo}, {Isaacson}, {J{\'o}zsa}, {Jonas}, {Joubert}, {Julie}, {Kapp},
  {Kenyon}, {Kotz{\'e}}, {Kriek}, {Kriel}, {Krishnan}, {Lehmensiek},
  {Liebenberg}, {Lord}, {Lunsky}, {Madisa}, {Magnus}, {Mahgoub}, {Makhaba},
  {Makhathini}, {Malan}, {Manley}, {Marais}, {Martens}, {Mauch}, {Merry},
  {Millenaar}, {Mnyandu}, {Mokone}, {Monama}, {Mphego}, {New}, {Ngcebetsha},
  {Ngoasheng}, {Ockards}, {Oozeer}, {Otto}, {Passmoor}, {Patel}, {Peens-Hough},
  {Perkins}, {Ramaila}, {Ramanujam}, {Ramudzuli}, {Ratcliffe}, {Robyntjies},
  {Salie}, {Sambu}, {Schollar}, {Schwardt}, {Schwartz}, {Serylak}, {Siebrits},
  {Sirothia}, {Slabber}, {Smirnov}, {Sofeya}, {Taljaard}, {Tasse}, {Tiplady},
  {Toruvanda}, {Twum}, {van Balla}, {van der Byl}, {van der Merwe}, {Van
  Tonder}, {Van Wyk}, {Venter}, {Venter}, {Wallace}, {Welz}, {Williams}, \&
  {Xaia}}]{Heywood2022}
{Heywood}, I., {Rammala}, I., {Camilo}, F., {et~al.} 2022, \apj, 925, 165,
  \dodoi{10.3847/1538-4357/ac449a}

\bibitem[{{Hobbs} {et~al.}(2020){Hobbs}, {Manchester}, {Dunning}, {Jameson},
  {Roberts}, {George}, {Green}, {Tuthill}, {Toomey}, {Kaczmarek}, {Mader},
  {Marquarding}, {Ahmed}, {Amy}, {Bailes}, {Beresford}, {Bhat}, {Bock},
  {Bourne}, {Bowen}, {Brothers}, {Cameron}, {Carretti}, {Carter}, {Castillo},
  {Chekkala}, {Cheng}, {Chung}, {Craig}, {Dai}, {Dawson}, {Dempsey}, {Doherty},
  {Dong}, {Edwards}, {Ergesh}, {Gao}, {Han}, {Hayman}, {Indermuehle},
  {Jeganathan}, {Johnston}, {Kanoniuk}, {Kesteven}, {Kramer}, {Leach},
  {Mcintyre}, {Moss}, {Os{\l}owski}, {Phillips}, {Pope}, {Preisig}, {Price},
  {Reeves}, {Reilly}, {Reynolds}, {Robishaw}, {Roush}, {Ruckley}, {Sadler},
  {Sarkissian}, {Severs}, {Shannon}, {Smart}, {Smith}, {Smith}, {Sobey},
  {Staveley-Smith}, {Tzioumis}, {van Straten}, {Wang}, {Wen}, \&
  {Whiting}}]{Hobbs2020}
{Hobbs}, G., {Manchester}, R.~N., {Dunning}, A., {et~al.} 2020, \pasa, 37,
  e012, \dodoi{10.1017/pasa.2020.2}

\bibitem[{{Hobbs} {et~al.}(2006){Hobbs}, {Edwards}, \&
  {Manchester}}]{Hobbs2006}
{Hobbs}, G.~B., {Edwards}, R.~T., \& {Manchester}, R.~N. 2006, \mnras, 369,
  655, \dodoi{10.1111/j.1365-2966.2006.10302.x}

\bibitem[{{Hotan} {et~al.}(2004){Hotan}, {van Straten}, \&
  {Manchester}}]{Hotan2004}
{Hotan}, A.~W., {van Straten}, W., \& {Manchester}, R.~N. 2004, \pasa, 21, 302,
  \dodoi{10.1071/AS04022}

\bibitem[{{Jankowski} {et~al.}(2018){Jankowski}, {van Straten}, {Keane},
  {Bailes}, {Barr}, {Johnston}, \& {Kerr}}]{jvk+18}
{Jankowski}, F., {van Straten}, W., {Keane}, E.~F., {et~al.} 2018, \mnras, 473,
  4436, \dodoi{10.1093/mnras/stx2476}

\bibitem[{{Johnston} {et~al.}(2006){Johnston}, {Kramer}, {Lorimer}, {Lyne},
  {McLaughlin}, {Klein}, \& {Manchester}}]{jkl+06}
{Johnston}, S., {Kramer}, M., {Lorimer}, D.~R., {et~al.} 2006, \mnras, 373, L6,
  \dodoi{10.1111/j.1745-3933.2006.00232.x}

\bibitem[{{Lazarevi{\'c}} {et~al.}(2023){Lazarevi{\'c}}, {Filipovi{\'c}},
  {Dai}, {Kothes}, {Ahmad}, {Alsaberi}, {Balzan}, {Barnes}, {Cotton},
  {Edwards}, {Gordon}, {Haberl}, {Hopkins}, {Koribalski}, {Leahy}, {Maitra},
  {Mi{\'c}i{\'c}}, {Rowell}, {Sasaki}, {Tothill}, {Umana}, \&
  {Velovi{\'c}}}]{Lazarevic2023}
{Lazarevi{\'c}}, S., {Filipovi{\'c}}, M.~D., {Dai}, S., {et~al.} 2023, arXiv
  e-prints, arXiv:2312.06961, \dodoi{10.48550/arXiv.2312.06961}

\bibitem[{{Lower} {et~al.}(2020){Lower}, {Shannon}, {Johnston}, \&
  {Bailes}}]{Lower2020}
{Lower}, M.~E., {Shannon}, R.~M., {Johnston}, S., \& {Bailes}, M. 2020, \apjl,
  896, L37, \dodoi{10.3847/2041-8213/ab9898}

\bibitem[{{Morello} {et~al.}(2019){Morello}, {Barr}, {Cooper}, {Bailes},
  {Bates}, {Bhat}, {Burgay}, {Burke-Spolaor}, {Cameron}, {Champion}, {Eatough},
  {Flynn}, {Jameson}, {Johnston}, {Keith}, {Keane}, {Kramer}, {Levin}, {Ng},
  {Petroff}, {Possenti}, {Stappers}, {van Straten}, \& {Tiburzi}}]{Morello2019}
{Morello}, V., {Barr}, E.~D., {Cooper}, S., {et~al.} 2019, \mnras, 483, 3673,
  \dodoi{10.1093/mnras/sty3328}

\bibitem[{{O'Leary} {et~al.}(2015){O'Leary}, {Kistler}, {Kerr}, \&
  {Dexter}}]{OLeary2015}
{O'Leary}, R.~M., {Kistler}, M.~D., {Kerr}, M., \& {Dexter}, J. 2015, arXiv
  e-prints, arXiv:1504.02477, \dodoi{10.48550/arXiv.1504.02477}

\bibitem[{{Pfahl} \& {Loeb}(2004)}]{Pfahl2004}
{Pfahl}, E., \& {Loeb}, A. 2004, \apj, 615, 253, \dodoi{10.1086/423975}

\bibitem[{{Ploeg} {et~al.}(2020){Ploeg}, {Gordon}, {Crocker}, \&
  {Macias}}]{ploeg20}
{Ploeg}, H., {Gordon}, C., {Crocker}, R., \& {Macias}, O. 2020, \jcap, 2020,
  035, \dodoi{10.1088/1475-7516/2020/12/035}

\bibitem[{{Ransom}(2011)}]{Ransom2011}
{Ransom}, S. 2011, {PRESTO: PulsaR Exploration and Search TOolkit},
  Astrophysics Source Code Library, record ascl:1107.017.
\newblock \doeprint{1107.017}

\bibitem[{{Rea} {et~al.}(2013){Rea}, {Esposito}, {Pons}, {Turolla}, {Torres},
  {Israel}, {Possenti}, {Burgay}, {Vigan{\`o}}, {Papitto}, {Perna}, {Stella},
  {Ponti}, {Baganoff}, {Haggard}, {Camero-Arranz}, {Zane}, {Minter},
  {Mereghetti}, {Tiengo}, {Sch{\"o}del}, {Feroci}, {Mignani}, \&
  {G{\"o}tz}}]{Rea2013}
{Rea}, N., {Esposito}, P., {Pons}, J.~A., {et~al.} 2013, \apjl, 775, L34,
  \dodoi{10.1088/2041-8205/775/2/L34}

\bibitem[{{Schnitzeler} {et~al.}(2016){Schnitzeler}, {Eatough}, {Ferri{\`e}re},
  {Kramer}, {Lee}, {Noutsos}, \& {Shannon}}]{sef+16}
{Schnitzeler}, D.~H.~F.~M., {Eatough}, R.~P., {Ferri{\`e}re}, K., {et~al.}
  2016, \mnras, 459, 3005, \dodoi{10.1093/mnras/stw841}

\bibitem[{{Shannon} \& {Johnston}(2013)}]{sj13}
{Shannon}, R.~M., \& {Johnston}, S. 2013, \mnras, 435, L29,
  \dodoi{10.1093/mnrasl/slt088}

\bibitem[{{Smith} {et~al.}(2023){Smith}, {Abdollahi}, {Ajello}, {Bailes},
  {Baldini}, {Ballet}, {Baring}, {Bassa}, {Gonzalez}, {Bellazzini}, {Berretta},
  {Bhattacharyya}, {Bissaldi}, {Bonino}, {Bottacini}, {Bregeon}, {Bruel},
  {Burgay}, {Burnett}, {Cameron}, {Camilo}, {Caputo}, {Caraveo}, {Cavazzuti},
  {Chiaro}, {Ciprini}, {Clark}, {Cognard}, {Corongiu}, {Orestano},
  {Crnogorcevic}, {Cuoco}, {Cutini}, {D'Ammando}, {de Angelis}, {DeCesar}, {De
  Gaetano}, {de Menezes}, {Deneva}, {de Palma}, {Di Lalla}, {Dirirsa}, {Di
  Venere}, {Dom{\'\i}nguez}, {Dumora}, {Fegan}, {Ferrara}, {Fiori},
  {Fleischhack}, {Flynn}, {Franckowiak}, {Freire}, {Fukazawa}, {Fusco},
  {Galanti}, {Gammaldi}, {Gargano}, {Gasparrini}, {Giacchino}, {Giglietto},
  {Giordano}, {Giroletti}, {Green}, {Grenier}, {Guillemot}, {Guiriec},
  {Gustafsson}, {Harding}, {Hays}, {Hewitt}, {Horan}, {Hou}, {Jankowski},
  {Johnson}, {Johnson}, {Johnston}, {Kataoka}, {Keith}, {Kerr}, {Kramer},
  {Kuss}, {Latronico}, {Lee}, {Li}, {Li}, {Limyansky}, {Longo}, {Loparco},
  {Lorusso}, {Lovellette}, {Lower}, {Lubrano}, {Lyne}, {Maan}, {Maldera},
  {Manchester}, {Manfreda}, {Marelli}, {Mart{\'\i}-Devesa}, {Mazziotta},
  {McEnery}, {Mereu}, {Michelson}, {Mickaliger}, {Mitthumsiri}, {Mizuno},
  {Moiseev}, {Monzani}, {Morselli}, {Negro}, {Nemmen}, {Nieder}, {Nuss},
  {Omodei}, {Orienti}, {Orlando}, {Ormes}, {Palatiello}, {Paneque},
  {Panzarini}, {Parthasarathy}, {Persic}, {Pesce-Rollins}, {Pillera}, {Poon},
  {Porter}, {Possenti}, {Principe}, {Rain{\`o}}, {Rando}, {Ransom}, {Ray},
  {Razzano}, {Razzaque}, {Reimer}, {Reimer}, {Renault-Tinacci}, {Romani},
  {S{\'a}nchez-Conde}, {Parkinson}, {Scotton}, {Serini}, {Sgr{\`o}}, {Shannon},
  {Sharma}, {Shen}, {Siskind}, {Spandre}, {Spinelli}, {Stappers}, {Stephens},
  {Suson}, {Tabassum}, {Tajima}, {Tak}, {Theureau}, {Thompson}, {Tibolla},
  {Torres}, {Valverde}, {Venter}, {Wadiasingh}, {Wang}, {Wang}, {Wang},
  {Weltevrede}, {Wood}, {Yan}, {Zaharijas}, {Zhang}, \& {Zhu}}]{Smith2023}
{Smith}, D.~A., {Abdollahi}, S., {Ajello}, M., {et~al.} 2023, \apj, 958, 191,
  \dodoi{10.3847/1538-4357/acee67}

\bibitem[{{Spiewak} {et~al.}(2022){Spiewak}, {Bailes}, {Miles},
  {Parthasarathy}, {Reardon}, {Shamohammadi}, {Shannon}, {Bhat}, {Buchner},
  {Cameron}, {Camilo}, {Geyer}, {Johnston}, {Karastergiou}, {Keith}, {Kramer},
  {Serylak}, {van Straten}, {Theureau}, \& {Venkatraman Krishnan}}]{spiewak22}
{Spiewak}, R., {Bailes}, M., {Miles}, M.~T., {et~al.} 2022, \pasa, 39, e027,
  \dodoi{10.1017/pasa.2022.19}

\bibitem[{{Thomas} {et~al.}(2020){Thomas}, {Pfrommer}, \&
  {En{\ss}lin}}]{Thomas2020}
{Thomas}, T., {Pfrommer}, C., \& {En{\ss}lin}, T. 2020, \apjl, 890, L18,
  \dodoi{10.3847/2041-8213/ab7237}

\bibitem[{{Uchida} {et~al.}(1992){Uchida}, {Morris}, \& {Yusef-Zadeh}}]{umy92}
{Uchida}, K., {Morris}, M., \& {Yusef-Zadeh}, F. 1992, \aj, 104, 1533,
  \dodoi{10.1086/116337}

\bibitem[{{van Straten} \& {Bailes}(2011)}]{vanStraten2011}
{van Straten}, W., \& {Bailes}, M. 2011, \pasa, 28, 1, \dodoi{10.1071/AS10021}

\bibitem[{{van Straten} {et~al.}(2012){van Straten}, {Demorest}, \&
  {Oslowski}}]{vanStraten2012}
{van Straten}, W., {Demorest}, P., \& {Oslowski}, S. 2012, Astronomical
  Research and Technology, 9, 237, \dodoi{10.48550/arXiv.1205.6276}

\bibitem[{{Yao} {et~al.}(2017){Yao}, {Manchester}, \& {Wang}}]{Yao2017}
{Yao}, J.~M., {Manchester}, R.~N., \& {Wang}, N. 2017, \apj, 835, 29,
  \dodoi{10.3847/1538-4357/835/1/29}

\bibitem[{{Yusef-Zadeh} {et~al.}(2022){Yusef-Zadeh}, {Arendt}, {Wardle},
  {Heywood}, \& {Cotton}}]{Yusef-Zadeh2022}
{Yusef-Zadeh}, F., {Arendt}, R.~G., {Wardle}, M., {Heywood}, I., \& {Cotton},
  W. 2022, \mnras, 517, 294, \dodoi{10.1093/mnras/stac2415}

\bibitem[{{Yusef-Zadeh} \& {Wardle}(2019)}]{Yusef-Zadeh2019}
{Yusef-Zadeh}, F., \& {Wardle}, M. 2019, \mnras, 490, L1,
  \dodoi{10.1093/mnrasl/slz134}

\bibitem[{{Yusef-Zadeh} {et~al.}(1997){Yusef-Zadeh}, {Wardle}, \&
  {Parastaran}}]{Yusef-Zadeh1997}
{Yusef-Zadeh}, F., {Wardle}, M., \& {Parastaran}, P. 1997, \apjl, 475, L119,
  \dodoi{10.1086/310484}

\bibitem[{{Yusef-Zadeh} {et~al.}(2024){Yusef-Zadeh}, {Zhao}, {Arendt},
  {Wardle}, {Heinke}, {Royster}, {Lang}, \& {Michail}}]{Yusef-Zadeh2024}
{Yusef-Zadeh}, F., {Zhao}, J.-H., {Arendt}, R., {et~al.} 2024, \mnras, 530,
  254, \dodoi{10.1093/mnras/stae549}

\end{thebibliography}
\bibliographystyle{aasjournal}



\end{document}